\documentclass[pra,twocolumn,superscriptaddress]{revtex4}
\usepackage{graphicx}
\usepackage{amsmath, amssymb}
\usepackage{bm,color}
\usepackage{amsthm}
\usepackage{braket,mathrsfs}
\usepackage{CJK}

\def\be{\begin{equation}}
\def\ee{\end{equation}}
\def\ba{\begin{eqnarray}}
\def\ea{\end{eqnarray}}
\def\la{\langle}
\def\ra{\rangle}
\def\arg{\,\mathrm{arg}\,}

\begin{document}
\begin{CJK}{UTF8}{gbsn}
	\title{Ehrenfest Breakdown  of the Mean-field Dynamics of Bose Gases}
	\author{Xizhi Han(韩希之)}
       \affiliation{International Center for Quantum Materials, School of Physics, Peking University, Beijing 100871, China}
    \author{Biao Wu(吴飙)}   \email{wubiao@pku.edu.cn}
      \affiliation{International Center for Quantum Materials, School of Physics, Peking University, Beijing 100871, China}
      \affiliation{Collaborative Innovation Center of Quantum Matter, Beijing 100871, China}
      \affiliation{Wilczek Quantum Center, College of Science, Zhejiang University of Technology,  Hangzhou 310014, China}
      
	\begin{abstract}
		The mean-field dynamics of a Bose gas is shown to break down at time 
		$\tau_h = (c_1/\gamma) \ln N$ where $\gamma$ is the 
		Lyapunov  exponent of the mean-field  theory, $N$ is the number of bosons, 
		and $c_1$ is a system-dependent constant. The breakdown time $\tau_h$ is essentially 
		the Ehrenfest time that characterizes the breakdown of the correspondence 
		between classical  and quantum dynamics. This breakdown  can be well 
		described by a quantum fidelity defined for one-particle reduced density matrices.  
		Our results are obtained with the formalism in particle-number 
		phase space and are illustrated with a triple-well model. The logarithmic 
		quantum-classical correspondence
		time may be verified experimentally with  Bose-Einstein condensates.
	\end{abstract}
	
	\date{\today}
	\maketitle
\end{CJK}
\section{Introduction}
The nonlinear Gross-Pitaevskii equation (GPE), as a mean-field theory, 
has been the dominant tool in describing the dynamics of Bose-Einstein condensates  
(BECs) in ultracold atomic gases~\cite{Dalfovo1999RMP,Yukalov2004}.  However, we face a quandary when the mean-field dynamics
of  a BEC becomes dynamically unstable or chaotic~\cite{DynamicalWu,Smerzi2002PRL,BECFidelity, FidelityBEC, KickedBEC, Thommen2003PRL,Penna2003PRL}: on one hand, 
one may regard this instability as an unphysical artifact resulted 
from the mean-field approximation, since
the exact dynamics of a BEC is governed by the many-body Schr\"odinger equation, which 
is linear and thus does not allow chaos; on the other hand, 
the dynamical instability was observed in experiments~\cite{Burger2001PRL,Niu2002PRL,Fallani2004PRL,SChu2005PRL,You2005PRL,InstabilityBEC} and it has been 
proved with  mathematical rigor that the GPE  describes correctly not only 
the ground state but also the dynamics of a BEC in the large $N$ 
limit ($N$ is the number of bosons)~\cite{Lieb2000PRA,DeriveGPE}.

Our aim in this work is to resolve this fundamental dilemma.  
Our study shows that the mean-field theory (the GPE) is only valid up to time 
\be
\tau_h = \frac{c_1}{\gamma} \ln N + o(\ln N), \label{eq:timescale}
\ee
where $\gamma$ is the 
Lyapunov exponent of the mean-field  dynamics and $c_1$ is a 
constant that depends only on systems.  With this time scale, the dilemma is resolved:
on one hand, in the large $N$ limit ($N\rightarrow\infty$), 
$\tau_h$ goes to infinity and thus the GPE is always valid just as proved rigorously 
in Ref.~\cite{DeriveGPE}; on the other hand, the time $\tau_h$ increases 
with $N$ only logarithmically and it is not a long time for a typical BEC experiment. 
For example, for the system studied in Ref.~\cite{DynamicalWu}, the Lyapunov time 
$\tau_\gamma=1/\gamma\sim 1$ ms.  As the number of atoms in a BEC prepared
in a typical experiment is around $10^4$,  we have  $\tau_h\sim 10$ ms. 
As a result, the dynamical instability or the breakdown of the mean-field dynamics 
can be easily observed in a typical experiment as reported in Ref.~\cite{Fallani2004PRL}.

This time scale $\tau_h$ is essentially the Ehrenfest time, which is the time that
the correspondence between the classical and quantum dynamics breaks down~\cite{Berman1978,Ehrenfest}. 
The usual Ehrenfest time $\tau_{\mathrm{Eh}}= (c_1 / \gamma) \ln(A/\hbar)$, where 
$\gamma$ is the Lyapunov exponent of the classical motion and $A$ is a typical 
action~\cite{Ehrenfest}. 
The similarity is due to that the GPE can be regarded as a classical equation 
in the large $N$ limit~\cite{Yaffe}.  Therefore, our result paves a way to experimental 
investigation of a fundamental relation in the quantum-classical correspondence --- the logarithmic behavior of the Ehrenfest time --- as $N$ can 
be varied in experiments. 

We cast the quantum dynamics onto the particle-number 
phase space (PNPS), which is a rearrangement of Fock states. In this phase space, for a nearly coherent state and in the large $N$ limit,
quantum many-body dynamics is equivalent to an ensemble of mean-field dynamics. 
When the mean-field motion is regular, mean-field trajectories will stay together and 
the Bose gas remains coherent.  If the mean-field motion is unstable or chaotic, mean-field 
trajectories will separate soon from each other exponentially, leading to decoherence of Bose gas 
and breakdown of the mean-field theory.  So, there are two distinct types of quantum 
dynamics, whose difference can be characterized by the quantum fidelity for one-particle 
reduced density matrices. 

We investigate the Ehrenfest breakdown numerically in the system of a BEC 
in a triple-well potential~\cite{triplewell, triplewell2, Penna2003PRE,Liu2007PRA,GuoWu}, 
which may be the simplest BEC model that embraces chaotic mean-field dynamics.  With this model, 
we verify numerically the Ehrenfest time and show that our quantum fidelity 
can well capture the characteristics of two different types of quantum dynamics. 

The  mean-field instability or breakdown has been discussed in literature~\cite{BECFidelity, FidelityBEC, KickedBEC, DynamicalWu, TunnelWu, Zurek, Anglin2001, ChristophPRA, ChristophPRL,WaveChaosPRA}. However,  
a general and explicit relation between mean-field chaos, number of particles and breakdown time is still lacking. And in PNPS not only such breakdown can be understood intuitively and quantitatively, but the significance of a local phase structure is also apparent, distortion of which leads to decoherence. 


\section{Particle-number phase space}
In Ref.~\cite{Yaffe}, it is shown that many quantum systems become classical in
the large $N$ limit.  A dilute Bose gas belongs to this class of quantum systems: 
its dynamics becomes classical and it is well described by the mean-field GPE in 
the large $N$ limit.  In this section, we introduce PNPS, 
where this quantum-classical correspondence in the large $N$ limit becomes transparent.

\subsection{Definition}
Any quantum state $\ket{\Psi}$ of a system of $N$ identical bosons with $M$ single-particle 
states can be regarded as a wavefunction $\varphi(\bm{x})$ over an $(M - 1)$-dimensional lattice space, which we call particle number phase space (PNPS), via
\be
\varphi(\bm{x}) \equiv \bra{0}\prod_{i = 1}^M \frac{\hat{a}_i^{N x_i}}{\sqrt{(N x_i)!}} \ket{\Psi}\,,
\ee
where $x_i$'s are entries of the $M$-dimensional vector $\bm{x}$, $N x_i \in \{0, \ldots, N\}$ for $1 \leq i \leq M$ and $\sum_i x_i = 1$. And $\hat{a}_i^\dagger$ and $\hat{a}_i$ are the creation and annihilation operators for the $i$-th single-particle state, with $[\hat{a}_i, \hat{a}^\dagger_j] = \delta_{ij}$ and $\hat{n}_i \equiv \hat{a}_i^\dagger \hat{a}_i$.
The continuous limit of PNPS is a hyperplane in $[0, 1]^M$ (defined by constraint $\sum_{i = 1}^M x_i = 1$), where we can define (for $i$ from $1$ to $M$)
	\ba
		&& \la x_i \ra \equiv \int \mathrm{d}\bm{x}\,x_i |\varphi(\bm{x})|^2 \\
		&& \la (\Delta x_i)^2 \ra \equiv \int \mathrm{d}\bm{x}\,(x_i - \la x_i \ra)^2 |\varphi(\bm{x})|^2 
	\ea
to characterize the average position and spread of the distribution $|\varphi(\bm{x})|^2$ over PNPS, given $\ket{\Psi}$ normalized. Of course for any finite $N$, the integral should be interpreted 
as summations over all $\bm{x}$ in PNPS.

As an example of our particular interest, we examine an SU($M$) coherent state $|\Psi\ra_c$ in PNPS:
	\be
		\ket{\Psi}_c \equiv \frac{1}{\sqrt{N!}} \left(\sum_{i = 1}^M \psi_i a_i^\dagger \right)^N\ket{0}\,, \label{eq:coherent}
	\ee
	where $\sum_i |\psi_i|^2 = 1$. In such case, we say $\ket{\psi}$ (an $M$-dimensional vector with $\psi_i$ as its entries) is the mean-field state of the SU($M$) coherent state $\ket{\Psi}_c$. It is straightforward to show for this coherent state $\ket{\Psi}_c$
	\be
		\la x_i \ra = |\psi_i|^2\,,\quad \la (\Delta x_i)^2 \ra = |\psi_i|^2 (1 - |\psi_i|^2) / N\,,\label{eq:width}
	\ee
	which indicate that the coherent state $\ket{\Psi}_c$ corresponds to a localized distribution $|\varphi(\bm{x})|^2$ in PNPS that peaks around $(|\psi_1|^2, |\psi_2|^2,\cdots, |\psi_M|^2)$ with a vanishing spread at large $N$. 
	
	And the wavefunction $\varphi(\bm{x})$ in PNPS has a phase structure.  For any $\bm{x}$ and $\bm{y}$ in PNPS,  
	\be
		\arg \varphi(\bm{x}) - \arg \varphi(\bm{y}) = N \sum_{i = 1}^M (x_i - y_i)\arg \psi_i \,\,(\textrm{mod } 2 \pi)\,,  \label{eq:phase}
	\ee
	which shows a wavevector $\bm{k}$: $k_i = N \arg \psi_i \propto N$. This phase structure is important as it will give us an estimate of the time $\tau_h$ in our later discussion. It is worth noting 
that when $N \to \infty$, there is no limit of the wavefunction $\varphi(\bm{x})$ because its wavevector $\bm{k}$ diverges.

	Overall, we find that the coherent state corresponds to a single-peaked wavepacket with plane-wave phase structure in PNPS.  In the following, we shall discuss quantum dynamics in PNPS and its relation to the mean-field dynamics. Note that the formalism of PNPS 
was also used in other contexts~\cite{PennaPRA, PennaPRA2}, where  phase 
structure and dynamics, however, were not discussed.

\subsection{Dynamics}
	Consider a quite general Hamiltonian of a Bose gas
	\be
		 \hat{\mathcal{H}} = \sum_{i, j = 1}^M\Big\{ H^{0}_{ij} \hat{a}_i^\dagger \hat{a}_j + \frac{U_{ij}}{N} \hat{a}_i^\dagger \hat{a}_j^\dagger \hat{a}_j \hat{a}_i\Big\}\,, \label{eq:Hamiltonian}
	\ee
where  $H_{ij}^0=H_{ji}^{0*}$ and $U_{ij}=U_{ji}$. 
Corresponding to the Schr\"{o}dinger equation $i \partial_t \ket{\Psi} = \hat{\mathcal{H}}\ket{\Psi}$, 
there is an equation of motion (EOM) for $\varphi(\bm{x}; t)$ in PNPS (Eq.\,(\ref{schrodinger}) in the Appendix). We are especially interested in the dynamics of a nearly coherent 
state $\varphi(\bm{x}; t)$, which satisfies the following two conditions:

(\textit{i})~the distribution $|\varphi(\bm{x}; t)|^2$ is localized such that $1 / N \ll \sqrt{\la (\Delta x_i)^2 \ra} \ll 1$ for all $i = 1, 2, \ldots, M$; 

(\textit{ii})~a local wavevector $\bm{k}(\bm{x}; t) \equiv \nabla_{\bm{x}} \arg \varphi(\bm{x}; t)$ exists in PNPS and varies insignificantly over a scale of $1 / N$, i.e., $|\partial_{x_i} k_j| \ll N$ for all $i, j = 1, 2, \ldots, M$.

With these two conditions and  keeping only finite terms in the large $N$ limit, an approximate (to $\mathcal{O}(1)$) EOM for $\varphi(\bm{x}; t)$ in PNPS can be 
derived (see Eq.\,(\ref{approx}) in the Appendix). 
Mathematically, there are $\delta$-function solutions to this EOM (Eq.\,(\ref{approx})): 
	\be
	\varphi(\bm{x}; t) = \exp[i \alpha(t)]\prod_{i = 1}^M \delta(x_i - x^0_i(t)) \exp [i k^0_i(t) x_i]\,.\label{eq:delta}
	\ee
In these $\delta$-function solutions, $x^0_i(t)$, $k^0_i(t)$ satisfy the following equation
	\be
    		i \partial_t \hat{\rho} = [\hat{\mathcal{H}}_{\mathrm{MF}}, \hat{\rho}]\,, \label{eq:rhomotion}
    \ee
    where $\mathcal{H}_{\mathrm{MF}, ij}(t) \equiv H^{0}_{ij} + 2 U_{ij} \rho_{ij}(t)$ and     	
    \be 
    		\rho_{ij}(t) \equiv \sqrt{x^0_i(t) x^0_j(t)} \,\mathrm{e}^{ i (k^0_i(t) - k^0_j(t))/N}\,.
    	\ee
This is just the mean-field EOM for the one-particle reduced density matrix.

Conditions ({\it i}) and ({\it ii}) reflect our expectations of nearly coherent states (see Eqs.\,(\ref{eq:width})  and (\ref{eq:phase})). The existence of $\delta$-function solutions corresponds to the established result that for any time $t_0$, when $N \to \infty$, coherent states at $t = 0$ stay coherent when $t = t_0$~\cite{DeriveGPE}. 

The results above can be interpreted as follows: at large $N$, for any initial state satisfying the two conditions, its time evolution may be regarded as the superposition of mean-field dynamics of $\delta$-functions, since any function in PNPS can be decomposed into a superposition of a cloud of $\delta$-functions! 
This is similar to the quantum dynamics of 
a single-particle wavepacket in real space: it can be regarded as a cloud of 
classical particles and each of them follows the Newton's EOM. 

As the quantum-classical correspondence between a quantum wavepacket and a classical particle will break down at the Ehrenfest time, the correspondence between one state in PNPS and its mean-field description --- one $\delta$-function solution (see Eq.\,(\ref{eq:delta})) --- will also fail when the mean-field trajectories of the $\delta$-functions in the cloud diverge. 

The breakdown time $\tau_h$ can be estimated using a conventional strategy in quantum chaos as in Ref.~\cite{Ehrenfest}. Essentially, before the breakdown the wavepacket of nearly coherent states in PNPS expands in the form of $\exp \gamma t$, where $\gamma$ is the Lyapunov exponent of the mean-field dynamics. According to Eq.\,(\ref{eq:width}), for $t < \tau_h$,
	    \be
	    		\Delta(t) \equiv \sqrt{\frac{1}{M}\sum_{i = 1}^M \la(\Delta x_i)^2\ra(t)} \propto \frac{\mathrm{e}^{\gamma t}}{\sqrt{N}}\,. \label{eq:Delta}
	    \ee
 
	   And there is a consistent mean-field description only if local wavevectors across the wavepacket are almost equal, that is,
	    \be
	    		\kappa t N^{-\lambda} \Delta(t) \ll 1\,, \label{eq:rate}
	    \ee
	    where $\kappa N^{-\lambda}$ is the average rate of growth of curvature $\partial_{x_i} k_j$ and the $N$ dependence is written explicitly. Substituting  (\ref{eq:Delta}) into (\ref{eq:rate}),  we have 
	    \be 
	    \gamma t + \ln t + \ln \kappa \ll \big(\lambda + \frac{1}{2}\big) \ln N\,.\label{eq:check}
	    \ee 
	    
The Ehrenfest time $\tau_h$ in Eq.\,(\ref{eq:timescale}) is obtained with $c_1 = \lambda + \frac{1}{2}$, which is independent of $N$ or $\gamma$. Numerical verification of this relation will be presented later.

Note that it is well-known that the quantum-classical correspondence may last far beyond the Ehrenfest time (see, e.g., Ref.\,\cite{ZurekPRA}). Similarly, it is possible that the mean-field theory remains valid even after our first estimate $t = \tau_h$; this interesting and special topic will be left for future study.

\begin{figure}
 \includegraphics[scale = 0.5]{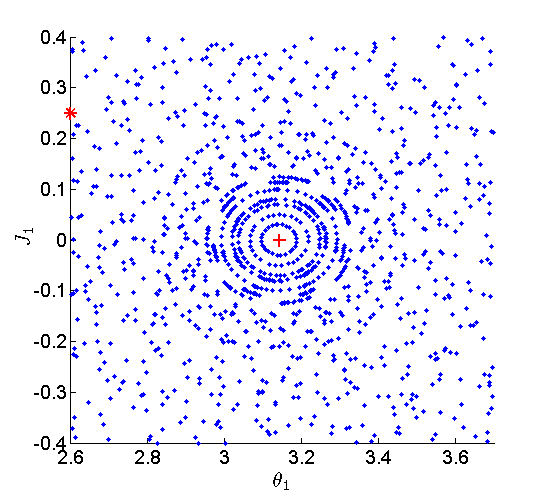}
 \caption{(color online) Poincar\'e section of the classical (mean-field) triple-well Hamiltonian with conjugate variables $(J_1, \theta_1)$ and $(J_2, \theta_2)$ at $\theta_2 = 0$, $\dot{\theta}_2 < 0$, $c = 1.25$, $E \approx 0.708$. `+' represents a state in the central regular region and `*' represents a state in the chaotic sea. }
 \label{poin}
\end{figure}

	\begin{figure}[t]
		    \includegraphics[scale = 0.3]{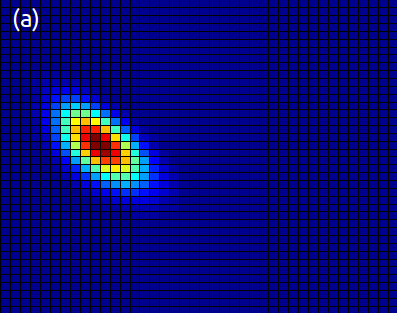}
		    \includegraphics[scale = 0.3]{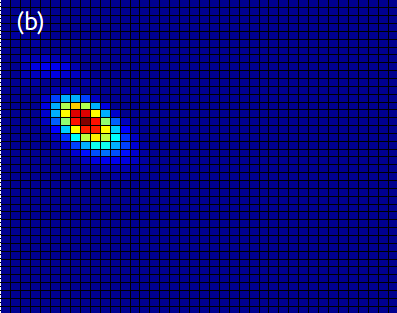}
		    \includegraphics[scale = 0.3]{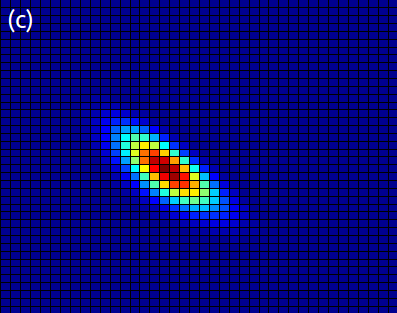}
		    \includegraphics[scale = 0.3]{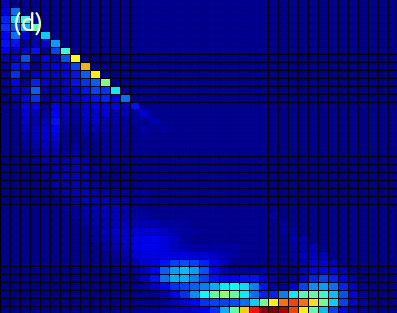}
		    \caption{(color online) Plot of $|\varphi(x_1, x_2, x_3; t)|^2$ for the quantum 
		    triple-well model with $N = 40$. Two axes are $x_1 \in [0, 1]$ and $x_2 \in [0, 1]$ ($x_3 = 1 - x_1 - x_2$). Red regions are of larger $|\varphi|^2$. (a) Initial state corresponding to 
		    the mean-field state denoted  by `+' in Fig.\,\ref{poin}; (b) the `+' state after evolving 
		    dynamically $t = 14.5$;  (c) initial state corresponding to the  `*' state in Fig.\,\ref{poin}; (d) the `*' state at $t = 14.5$.}          
		    \label{fig:schematic}
	    \end{figure}

\section{Example of Triple-well Model}
We now illustrate our results with an example. Consider a BEC in a ring-shaped triple-well potential~\cite{GuoWu}.  Under tight-binding approximation,  the second-quantized Hamiltonian is (as a specific case of Eq.\,(\ref{eq:Hamiltonian}))
	\be
		\hat{\mathcal{H}} = -\frac{1}{2}\sum_{1 \leq i, j \leq 3}^{i \neq j} \hat{a}_i^\dagger \hat{a}_j + \frac{c}{2 N} \sum_{i = 1}^3 \hat{a}_i^\dagger \hat{a}_i^\dagger \hat{a}_i \hat{a}_i\,,
	\ee 
	where $c$ is the on-site interaction strength. For this system $M = 3$.  Its corresponding nonlinear mean-field EOM is
	\be
		i\frac{\mathrm{d}}{\mathrm{d}t}\left(
		\begin{array}{c}
			\psi_1 \\
			\psi_2 \\
			\psi_3 \\
		\end{array}
		\right) = \left(
		\begin{array}{ccc}
			 c |\psi_1|^2 & -1 / 2 & -1 / 2 \\
			-1 / 2 & c |\psi_2|^2  & -1 / 2 \\
			-1 / 2 & -1 / 2 & c |\psi_3|^2
		\end{array}\right)
		\left(
		\begin{array}{c}
			\psi_1 \\
			\psi_2 \\
			\psi_3 \\
		\end{array}
		\right)\,.
		\label{eq:mf}
	\ee
Shown in Fig.\,\ref{poin} is a Poincar\'e section of the above mean-field dynamics, where
two kinds of motion are evident: the central regular region is surrounded by a chaotic sea. 
The  conjugate variables  used in plotting Fig.\,\ref{poin} are  $(J_1, \theta_1)$, $(J_2, \theta_2)$, which are defined as  $J_1 = |\psi_{1}|^2 - |\psi_{3}|^2$, $J_2 = |\psi_{3}|^2$, $\theta_1 = \arg \psi_{2} - \arg \psi_{1}$, $\theta_2 = 2 \arg \psi_{2} - \arg \psi_{1} - \arg \psi_{3}$. 
	
The quantum dynamics of this model can also be computed 	rather easily.  
The evolution of $|\varphi(\bm{x})|^2$ in PNPS is plotted in Fig.\,\ref{fig:schematic}, 
where two types of quantum dynamics are clearly observed. In Fig.\,\ref{fig:schematic} (a, b), 
an initial coherent state,  which is a gaussian-like wavepacket in PNPS, 
shows no significant expansion or distortion during dynamical evolution.  
In Fig.\,\ref{fig:schematic} (c, d),  the situation is drastically different: 
a similar-looking initial coherent state expands and becomes dramatically distorted after a certain time.
The difference is caused by the fact that the initial state in Fig.\,\ref{fig:schematic} (a)
corresponds to a mean-field state in the regular region in Fig.\,\ref{poin} while the one
in Fig.\,\ref{fig:schematic} (c) corresponds to a mean-field state in the chaotic region.  

It is obvious that the mean-field theory cannot describe the dramatic quantum dynamics 
shown in Fig.\,\ref{fig:schematic} (c, d). Such a failure or breakdown of the mean-field 
theory due to rapid decoherence has long been noticed in literature~\cite{Anglin2001,ChristophPRA, ChristophPRL,WaveChaosPRA}. In Ref.~\cite{ChristophPRL},  a remedy was tried 
unsuccessfully to bridge the gap between the mean-field theory and the exact quantum theory. 
In this work we have shown that there exists a general time scale $\tau_h$ in terms of Lyapunov exponent and number of bosons beyond which the mean-field theory fails.  In the following, we shall 
introduce a quantum fidelity to distinguish the two types of quantum
dynamics shown in Fig.\,\ref{fig:schematic} without using mean-field formalism, and confirm the time scale $\tau_h$ numerically. 

\subsection{Quantum Fidelity}
To quantify the loss of coherence in the quantum evolution as shown in Fig.\,\ref{fig:schematic} (d),
we introduce the following quantum fidelity ${\mathcal F}_q$ for one-particle reduced density matrix (RDM) $\hat{\rho}$ and $\hat{\chi}$:
\be
{\mathcal F}_q(\hat{\rho}, \hat{\chi}) \equiv \frac{1}{N^2} \mathrm{tr}\,\hat{\rho}^\dagger \hat{\chi}\,. \label{def:many}
\ee
For a quantum state $\ket{\Psi(t)}$,  its one-particle RDM  can be explicitly written as
\be \sum_{ij} \ket{i}\la \Psi(t) | \hat{a}_i^\dagger \hat{a}_j | \Psi(t) \ra\bra{j}\,.\ee
There are three reasons to use this quantum fidelity: 

1) Experimentally we are often interested in the one-particle RDM. 

2) It allows us to define coherence $\mathcal{C}$:
	\be \mathcal{C} (\hat{\rho}) \equiv {\mathcal F}_q(\hat{\rho}, \hat{\rho})\,,\ee
	where $\hat{\rho}$ is the one-particle RDM for $\ket{\Psi}$. The coherence $\mathcal{C}$
	can quantify how coherent the state $\ket{\Psi}$ is: $\mathcal{C} (\hat{\rho}) = 1$ if and only if $\ket{\Psi}$ is a coherent state as in Eq.\,(\ref{eq:coherent}).
	
3) It returns to the mean-field fidelity for 
coherent states, i.e., ${\mathcal F}_q(\hat{\rho}, \hat{\chi}) = {\mathcal F}_{\mathrm{mf}}(\psi, \phi) \equiv|\la \phi | \psi \ra|^2$ if $\hat{\rho}$, $\hat{\chi}$ are one-particle RDM for coherent states $\ket{\Psi}_c$ and $\ket{\Phi}_c$, and $\psi$, $\phi$ are mean-field states of $\ket{\Psi}_c$ and $\ket{\Phi}_c$ (see discussion under Eq.\,(\ref{eq:coherent})). Therefore, before the Ehrenfest breakdown ${\mathcal F}_q$ essentially captures mean-field characteristics, especially the Lyapunov exponent, which distinguishes regular and chaotic mean-field trajectories.

\subsection{Numerical Results}


		\begin{figure}[t]
		    \includegraphics[scale = 0.24]{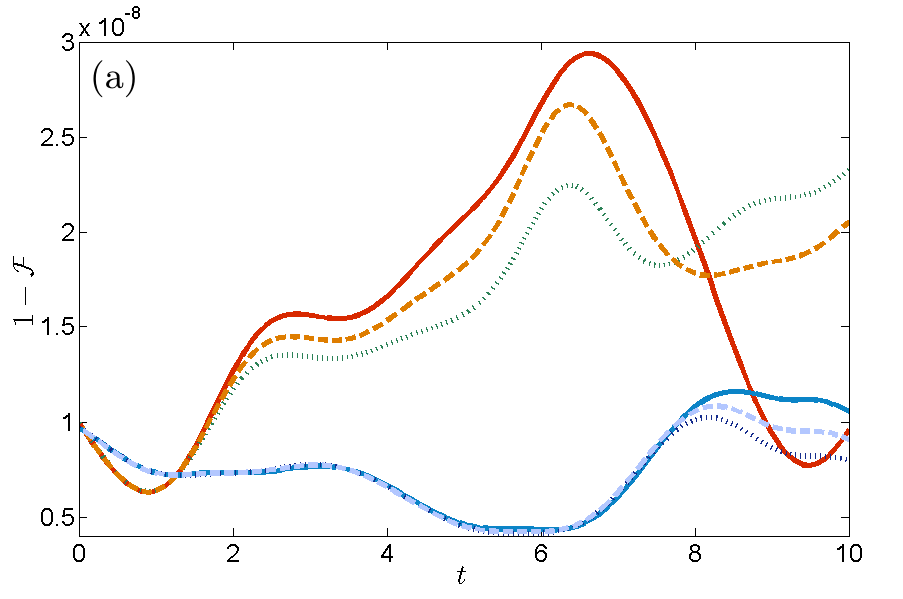}
		    \includegraphics[scale = 0.24]{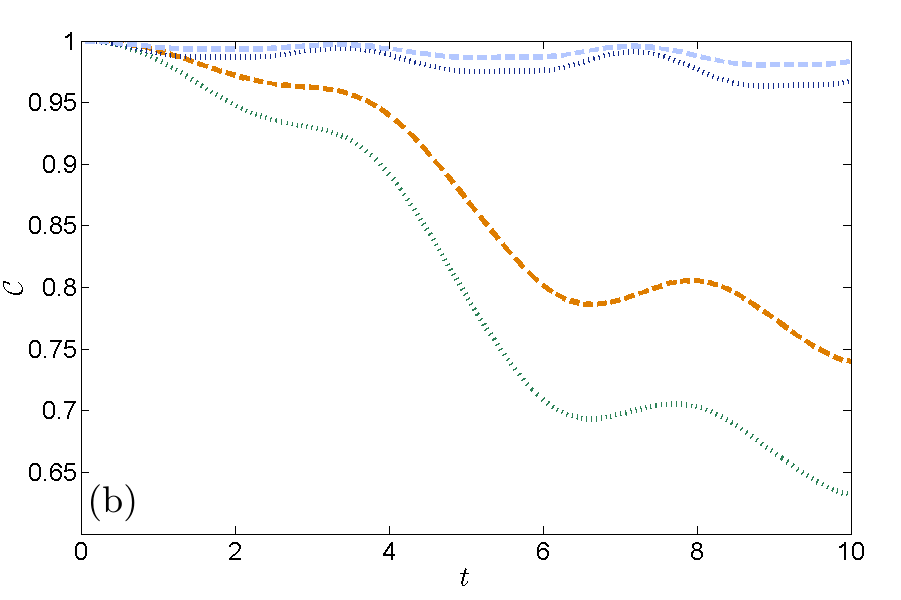}
		    \caption{(color online) (a) Quantum and mean-field fidelities. Solid lines are $1 - \mathcal{F}_\mathrm{mf}(\psi(t), \tilde{\psi}(t))$;  dashed lines are $1 - \mathcal{F}_q(\hat{\rho}(t), \hat{\tilde{\rho}}(t)) / \sqrt{\mathcal{F}_q(\hat{\rho}(t), \hat{\rho}(t)) \mathcal{F}_q(\hat{\tilde{\rho}}(t), \hat{\tilde{\rho}}(t))}$ for $N = 80$; dotted lines are $1 - \mathcal{F}_q(\hat{\rho}(t), \hat{\tilde{\rho}}(t)) / \sqrt{\mathcal{F}_q(\hat{\rho}(t), \hat{\rho}(t)) \mathcal{F}_q(\hat{\tilde{\rho}}(t), \hat{\tilde{\rho}}(t))}$ for $N = 40$. $\mathcal{F}_q$ is normalized to better show the correspondence. (b) Coherence $\mathcal{C}(\hat{\rho}(t))$. Curves show the decay of coherence of quantum many-body states  in (a). 
		    In the simulation, $c = 1.25$, $E \approx 0.708$, $\theta_1 = \pi$, $\theta_2 = 0$, $\hat{\rho}(t)$ and $\hat{\tilde{\rho}}(t)$ are the RDM of quantum states, whose corresponding mean-field states are $|\psi(t)\ra$ and $|\tilde{\psi}(t)\ra$, respectively. $\|\psi - \tilde{\psi}\|_{t = 0} \approx 10^{-4}$. The lower set of lines in (a) and the corresponding upper set of lines in (b) are for the integrable case $J_1 = 0$; the upper set in (a) and the corresponding lower set  in (b) are for the chaotic case $J_1 = 0.5$. }          
		    \label{fig:lyapunov}
	    \end{figure}

The numerical simulation aims at verifying our theoretical understanding as discussed: for a coherent initial state, at the beginning the mean-field dynamics agrees with the quantum evolution, producing even the same growth of discrepancy between states; however, long-time exponential growth is not allowed by quantum mechanics, so there exists an Ehrenfest time $\tau_h$ beyond which the mean-field and quantum correspondence fails. Such a failure is due to the decoherence of quantum states; the breakdown
time  $\tau_h$ is given in Eq.\,(\ref{eq:timescale}). In the following we provide numerical evidences for our theortical understanding.

We choose a coherent initial state $\ket{\Psi(t = 0)}_c$ with one-particle RDM $\hat{\rho}(t = 0)$, whose corresponding mean-field state is $\ket{\psi(t = 0)}$. Then we slightly perturb the mean-field state into $\ket{\tilde{\psi}(t = 0)}$, and generate the corresponding coherent state $\ket{\tilde{\Psi}(t = 0)}_c$ and RDM $\hat{\tilde{\rho}}(t = 0)$. Next we observe the evolution of quantum fidelity between these two states, which allows us to calculate the Lyapunov exponent. Of course, $\ket{\psi(t)}$ and $\ket{\tilde{\psi}(t)}$ evolve according to the mean-field equations Eq.\,(\ref{eq:mf}), $\ket{\Psi(t)}$ and $\ket{\tilde{\Psi}(t)}$ evolve according to the quantum Hamiltonian in Eq.\,(\ref{eq:Hamiltonian}), $\hat{\rho}(t)$ and $\hat{\tilde{\rho}}(t)$ are obtained from $\ket{\Psi(t)}$ and $\ket{\tilde{\Psi}(t)}$, respectively. $1 - {\mathcal F}_q(\hat{\rho}(t), \hat{\tilde{\rho}}(t))$ and $1 - {\mathcal F}_\mathrm{mf} (\psi(t), \tilde{\psi}(t))$ are shown in Fig.\,\ref{fig:lyapunov} (a), 
where we see that the mean-field fidelity ${\mathcal F}_{\rm mf}$ coincides with 
${\mathcal F}_{q}$ for small $t$, as expected. 

However, we also observe in Fig.\,\ref{fig:lyapunov} (a) that there is an Ehrenfest time $\tau_h$, when ${\mathcal F}_q$ and ${\mathcal F}_\mathrm{mf}$ start to visibly disagree. Cases for different $N$ and $\gamma$ are plotted in Fig.\,\ref{fig:lyapunov} (a), where we can see that as $N$ increases or $\gamma$ decreases, $\tau_h$ gets longer. This qualitatively agrees with the scaling of the Ehrenfest time. And in Fig.\,\ref{fig:lyapunov} (b), it is observed that although $\tau_h$ is different for different $N$ and $\gamma$, $\tau_h$ is approximately the time when the coherence ${\mathcal C}(\hat{\rho}(t))$ drops below 98\%. This confirms our understanding that the failure of correspondence between the mean-field and quantum descriptions is the result of decoherence of quantum states.

	    	\begin{figure}[t]
		    \includegraphics[scale=0.24]{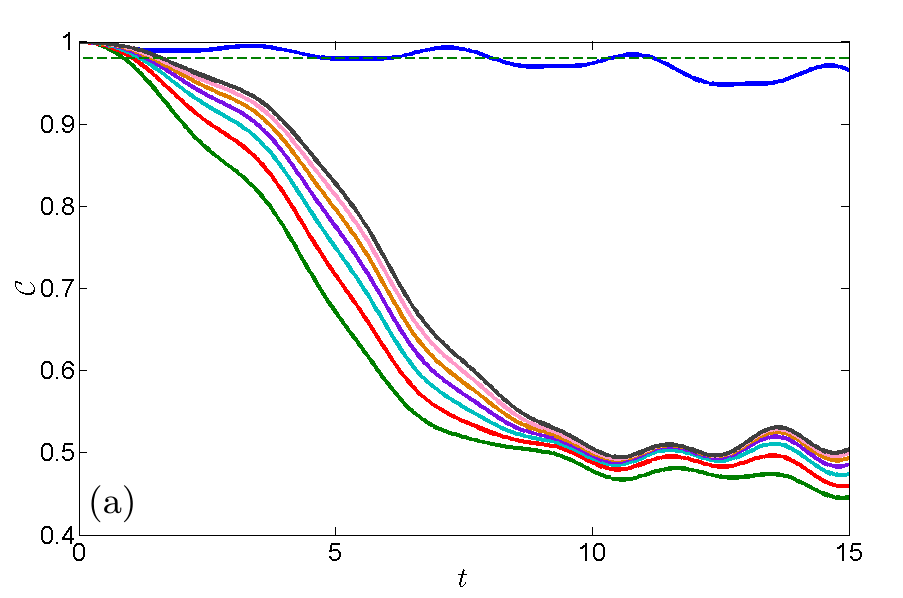}
		    \includegraphics[scale=0.24]{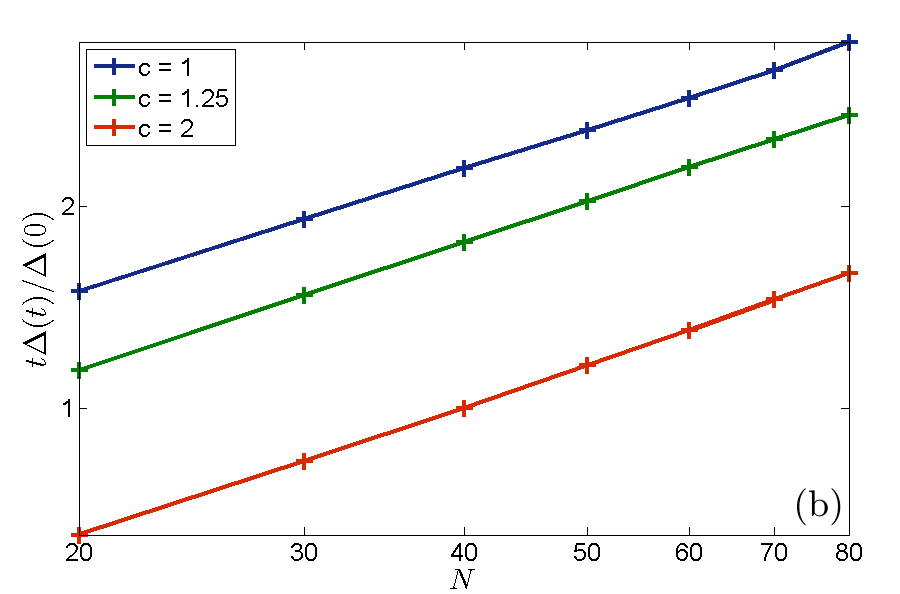}
		    \caption{(color online) (a) Time evolution of coherence $\mathcal{C}(\hat{\rho}(t))$  for the integrable $J_1 = 0$ and the chaotic $J_1 = 0.6$ trajectories. $c = 1.25$, $E \approx 0.708$, $\theta_1 = \pi$, $\theta_2 = 0$ and time step in simulation is $10^{-3}$. $\mathcal{C}$ of integrable $J_1 = 0$ (the solid line in the top) remains high while $\mathcal{C}$ of chaotic $J_1 = 0.6$ (other solid lines) drops quickly ($N = 20, 30, \ldots, 80$ from left to right). The dashed line is $\mathcal{C} = 98\%$. (b) Linear fitting of $\ln \Delta(t) / \Delta(0) + \ln t + \ln \kappa = c_1 \ln N$ for the chaotic initial state in (a) with $c = 1$, 1.25 and 2, $N$ from 20 to 80. Data points are calculated when $\mathcal C$ drops to $98\%$. The slope $c_1 \approx 0.6$ is found independent of $c$ or $\gamma$.  }          
		    \label{fig2}
	    \end{figure}
	    
Based on such understanding, we can quantitatively define the Ehrenfest time in this example as the time when the coherence ${\mathcal C}(\hat{\rho}(t))$ drops below 98\%. Examples of decay of ${\mathcal C}(\hat{\rho}(t))$ is illustrated in Fig.\,\ref{fig2} (a), where the Ehrenfest time $\tau_h$ is measured when ${\mathcal C}(\hat{\rho}(t))$ drops below the dashed line. By varying $N$ and $c$, we verify the relation Eq.\,(\ref{eq:check}), which leads to Eq.\,(\ref{eq:timescale}), in Fig.\,\ref{fig2} (b). A linear fitting between $\ln \Delta(t) / \Delta(0) + \ln t$ and $\ln N$ is found with a constant slope $\lambda + \frac{1}{2}\approx 0.6$ (see Eq.\,(\ref{eq:check})), suggesting $c_1 \approx 0.6$ in Eq.\,(\ref{eq:timescale}). Note here $\gamma t$ is replaced by $\ln \Delta(t) / \Delta(0)$ for numerical convenience.
	
\section{Conclusion}
	In sum we have answered an intriguing question --- when does mean-field approximation of a dilute Bose gas remain valid as the system evolves? Our answer is the mean-field dynamics breaks down at the Ehrenfest time $\tau_h=(c_1/\gamma)\ln N$.  The study is facilitated by introducing particle number phase space,  where one can see easily that the correspondence between many-body quantum dynamics and mean-field dynamics is similar to the usual quantum-classical correspondence. 
	
	 As $N$ can be varied 
in BEC experiments, it is now possible to experimentally measure the logarithmic behavior of the Ehrenfest time. One can compare physical observables in the experiment with their theoretical mean-field values, and measure the Ehrenfest time when their discrepancy exceeds a threshold. BECs with unstable or chaotic mean-field descriptions are suitable for such experiments; for example, spinor BECs~\cite{Chapman2005NPhys,YouLiPRA2005} may be a good candidate system. \\
		
This work is supported by the National Basic Research Program of China (2013CB921903, 2012CB921300) and
the National Natural Science Foundation of China (11274024, 11334001, 11429402).

\appendix
\section{Quantum EOM in PNPS and Its Mean-field Approximation}
For the Hamiltonian in Eq.\,(\ref{eq:Hamiltonian}), the Schr\"{o}dinger equation in PNPS reads ($\hbar = 1$)
	\ba
		 i \partial_t \varphi(\bm{x}; t) &=&  \sum_i H^{0}_{ii} N x_i \varphi(\bm{x}; t) \nonumber \\
		 & +& \sum_i U_{ii} N x_i (x_i - \epsilon) \varphi(\bm{x}; t) \nonumber\\ 
		 & +& \sum_{i \neq j} H^{0}_{ij} N \sqrt{x_i(x_j + \epsilon)} \varphi(\bm{x} + \epsilon \bm{e}^{ij}; t) \nonumber\\
		 & +& \sum_{i \neq j} U_{ij} N x_i x_j \varphi(\bm{x}; t)\,, \label{schrodinger}
	\ea
	where $\epsilon \equiv 1 / N$, $\bm{e}^{ij}$ is an $M$-dimensional vector $e^{ij}_k \equiv -\delta_{ik} + \delta_{jk}$, $k = 1, 2, \ldots, M$.
	
We are especially interested in the dynamics of a nearly coherent state. With conditions ({\it i}) and ({\it ii}) in Sect. II B and $N \to \infty$, Eq.\,(\ref{schrodinger}) becomes
	\ba
		 i \partial_t \varphi &= & \sum_i H^{0}_{ii} N x_i \varphi + \sum_i U_{ii} N x_i (x_i - \epsilon) \varphi\nonumber\\ 
		 & + &\sum_{i \neq j} H^{0}_{ij} N \sqrt{x_i x_j} \varphi \exp [i(k_j - k_i)\epsilon] \nonumber\\
		 & + &\sum_{i \neq j} H^{0}_{ij} \frac{1}{2} \sqrt{\frac{x_i}{x_j}} \varphi \exp[ i(k_j - k_i)\epsilon] \nonumber\\
		 & + &\sum_{i \neq j} H^{0}_{ij} \sqrt{x_i x_j} [(\partial_j - \partial_i) |\varphi|] \frac{\varphi}{|\varphi|}\exp[ i(k_j - k_i)\epsilon] \nonumber\\
		 & + &\sum_{i \neq j} U_{ij} N x_i x_j \varphi\, + o(1), \label{approx}
	\ea
	where $\partial_i \equiv \frac{\partial}{\partial {x_i}}$ and $k_i(\bm{x}; t)$ is the local wavevector of wavefunction $\varphi$ at $(\bm{x}; t)$, as discussed in condition ({\it ii}). The argument of all $k_i$ and $\varphi$ is $(\bm{x}; t)$ and omitted.  
	
Now we assume a $\delta$-function solution as in Eq.\,(\ref{eq:delta}).
By equalling the coefficients before $\delta$, $\partial_i \delta$ and the derivatives of coefficients before $\delta$ (which is necessary to reflect the plane-wave phase structure) on both sides, keeping finite terms in the large $N$ limit, we obtain
\ba
    \partial_t x^0_i &=& 2 \,\mathrm{Im} \sum_{j} H^0_{ij} \sqrt{x^0_i x^0_j} \exp i (k^0_j - k^0_i)\epsilon \label{eq:x}
    \ea
    \ba
    \partial_t k^0_i \epsilon &=&  - \mathrm{Re} \sum_{j}  H^0_{ij}\sqrt{\frac{x^0_j}{x^0_i}} \exp i (k^0_j - k^0_i) \epsilon\nonumber\\
      && - 2\,\mathrm{Re} \sum_{j} U_{ij} x^0_j\,, \label{eq:k}
    \ea
    where the argument $t$ of all $x^0_i$ and $k^0_i$ is omitted for brevity. Lengthy but straightforward calculations will verify that Eqs.\,(\ref{eq:x}) and (\ref{eq:k}) are equivalent to Eq.\,(\ref{eq:rhomotion}), which is same as the mean-field EOM for the one particle RDM $\hat{\rho}$.\\


\begin{thebibliography}{35}
\expandafter\ifx\csname natexlab\endcsname\relax\def\natexlab#1{#1}\fi
\expandafter\ifx\csname bibnamefont\endcsname\relax
  \def\bibnamefont#1{#1}\fi
\expandafter\ifx\csname bibfnamefont\endcsname\relax
  \def\bibfnamefont#1{#1}\fi
\expandafter\ifx\csname citenamefont\endcsname\relax
  \def\citenamefont#1{#1}\fi
\expandafter\ifx\csname url\endcsname\relax
  \def\url#1{\texttt{#1}}\fi
\expandafter\ifx\csname urlprefix\endcsname\relax\def\urlprefix{URL }\fi
\providecommand{\bibinfo}[2]{#2}
\providecommand{\eprint}[2][]{\url{#2}}

\bibitem[{\citenamefont{Dalfovo et~al.}(1999)\citenamefont{Dalfovo, Giorgini,
  Pitaevskii, and Stringari}}]{Dalfovo1999RMP}
\bibinfo{author}{\bibfnamefont{F.}~\bibnamefont{Dalfovo}},
  \bibinfo{author}{\bibfnamefont{S.}~\bibnamefont{Giorgini}},
  \bibinfo{author}{\bibfnamefont{L.~P.} \bibnamefont{Pitaevskii}},
  \bibnamefont{and}
  \bibinfo{author}{\bibfnamefont{S.}~\bibnamefont{Stringari}},
  \bibinfo{journal}{Rev. Mod. Phys.} \textbf{\bibinfo{volume}{71}},
  \bibinfo{pages}{463} (\bibinfo{year}{1999}).

\bibitem[{\citenamefont{Yukalov}(2004)}]{Yukalov2004}
\bibinfo{author}{\bibfnamefont{V.~I.} \bibnamefont{Yukalov}},
  \bibinfo{journal}{Laser Physics Letters} \textbf{\bibinfo{volume}{1}},
  \bibinfo{pages}{435} (\bibinfo{year}{2004}).

\bibitem[{\citenamefont{Wu and Niu}(2001)}]{DynamicalWu}
\bibinfo{author}{\bibfnamefont{B.}~\bibnamefont{Wu}} \bibnamefont{and}
  \bibinfo{author}{\bibfnamefont{Q.}~\bibnamefont{Niu}},
  \bibinfo{journal}{Phys. Rev. A} \textbf{\bibinfo{volume}{64}},
  \bibinfo{pages}{061603} (\bibinfo{year}{2001}).

\bibitem[{\citenamefont{Smerzi et~al.}(2002)\citenamefont{Smerzi, Trombettoni,
  Kevrekidis, and Bishop}}]{Smerzi2002PRL}
\bibinfo{author}{\bibfnamefont{A.}~\bibnamefont{Smerzi}},
  \bibinfo{author}{\bibfnamefont{A.}~\bibnamefont{Trombettoni}},
  \bibinfo{author}{\bibfnamefont{P.~G.} \bibnamefont{Kevrekidis}},
  \bibnamefont{and} \bibinfo{author}{\bibfnamefont{A.~R.}
  \bibnamefont{Bishop}}, \bibinfo{journal}{Phys. Rev. Lett.}
  \textbf{\bibinfo{volume}{89}}, \bibinfo{pages}{170402}
  (\bibinfo{year}{2002}).

\bibitem[{\citenamefont{Liu et~al.}(2006)\citenamefont{Liu, Wang, Zhang, Niu,
  and Li}}]{BECFidelity}
\bibinfo{author}{\bibfnamefont{J.}~\bibnamefont{Liu}},
  \bibinfo{author}{\bibfnamefont{W.}~\bibnamefont{Wang}},
  \bibinfo{author}{\bibfnamefont{C.}~\bibnamefont{Zhang}},
  \bibinfo{author}{\bibfnamefont{Q.}~\bibnamefont{Niu}}, \bibnamefont{and}
  \bibinfo{author}{\bibfnamefont{B.}~\bibnamefont{Li}},
  \bibinfo{journal}{Physics Letters A} \textbf{\bibinfo{volume}{353}},
  \bibinfo{pages}{216 } (\bibinfo{year}{2006}), ISSN \bibinfo{issn}{0375-9601}.

\bibitem[{\citenamefont{Manfredi and Hervieux}(2008)}]{FidelityBEC}
\bibinfo{author}{\bibfnamefont{G.}~\bibnamefont{Manfredi}} \bibnamefont{and}
  \bibinfo{author}{\bibfnamefont{P.-A.} \bibnamefont{Hervieux}},
  \bibinfo{journal}{Phys. Rev. Lett.} \textbf{\bibinfo{volume}{100}},
  \bibinfo{pages}{050405} (\bibinfo{year}{2008}).

\bibitem[{\citenamefont{Reslen et~al.}(2008)\citenamefont{Reslen, Creffield,
  and Monteiro}}]{KickedBEC}
\bibinfo{author}{\bibfnamefont{J.}~\bibnamefont{Reslen}},
  \bibinfo{author}{\bibfnamefont{C.~E.} \bibnamefont{Creffield}},
  \bibnamefont{and} \bibinfo{author}{\bibfnamefont{T.~S.}
  \bibnamefont{Monteiro}}, \bibinfo{journal}{Phys. Rev. A}
  \textbf{\bibinfo{volume}{77}}, \bibinfo{pages}{043621}
  (\bibinfo{year}{2008}).

\bibitem[{\citenamefont{Thommen et~al.}(2003)\citenamefont{Thommen, Garreau,
  and Zehnl\'e}}]{Thommen2003PRL}
\bibinfo{author}{\bibfnamefont{Q.}~\bibnamefont{Thommen}},
  \bibinfo{author}{\bibfnamefont{J.~C.} \bibnamefont{Garreau}},
  \bibnamefont{and} \bibinfo{author}{\bibfnamefont{V.}~\bibnamefont{Zehnl\'e}},
  \bibinfo{journal}{Phys. Rev. Lett.} \textbf{\bibinfo{volume}{91}},
  \bibinfo{pages}{210405} (\bibinfo{year}{2003}).

\bibitem[{\citenamefont{Buonsante et~al.}(2003)\citenamefont{Buonsante,
  Franzosi, and Penna}}]{Penna2003PRL}
\bibinfo{author}{\bibfnamefont{P.}~\bibnamefont{Buonsante}},
  \bibinfo{author}{\bibfnamefont{R.}~\bibnamefont{Franzosi}}, \bibnamefont{and}
  \bibinfo{author}{\bibfnamefont{V.}~\bibnamefont{Penna}},
  \bibinfo{journal}{Phys. Rev. Lett.} \textbf{\bibinfo{volume}{90}},
  \bibinfo{pages}{050404} (\bibinfo{year}{2003}).

\bibitem[{\citenamefont{Burger et~al.}(2001)\citenamefont{Burger, Cataliotti,
  Fort, Minardi, Inguscio, Chiofalo, and Tosi}}]{Burger2001PRL}
\bibinfo{author}{\bibfnamefont{S.}~\bibnamefont{Burger}},
  \bibinfo{author}{\bibfnamefont{F.~S.} \bibnamefont{Cataliotti}},
  \bibinfo{author}{\bibfnamefont{C.}~\bibnamefont{Fort}},
  \bibinfo{author}{\bibfnamefont{F.}~\bibnamefont{Minardi}},
  \bibinfo{author}{\bibfnamefont{M.}~\bibnamefont{Inguscio}},
  \bibinfo{author}{\bibfnamefont{M.~L.} \bibnamefont{Chiofalo}},
  \bibnamefont{and} \bibinfo{author}{\bibfnamefont{M.~P.} \bibnamefont{Tosi}},
  \bibinfo{journal}{Phys. Rev. Lett.} \textbf{\bibinfo{volume}{86}},
  \bibinfo{pages}{4447} (\bibinfo{year}{2001}).

\bibitem[{\citenamefont{Wu and Niu}(2002)}]{Niu2002PRL}
\bibinfo{author}{\bibfnamefont{B.}~\bibnamefont{Wu}} \bibnamefont{and}
  \bibinfo{author}{\bibfnamefont{Q.}~\bibnamefont{Niu}},
  \bibinfo{journal}{Phys. Rev. Lett.} \textbf{\bibinfo{volume}{89}},
  \bibinfo{pages}{088901} (\bibinfo{year}{2002}).

\bibitem[{\citenamefont{Fallani et~al.}(2004)\citenamefont{Fallani, Sarlo, Lye,
  Modugno, Saers, Fort, and Inguscio}}]{Fallani2004PRL}
\bibinfo{author}{\bibfnamefont{L.}~\bibnamefont{Fallani}},
  \bibinfo{author}{\bibfnamefont{L.~D.} \bibnamefont{Sarlo}},
  \bibinfo{author}{\bibfnamefont{J.}~\bibnamefont{Lye}},
  \bibinfo{author}{\bibfnamefont{M.}~\bibnamefont{Modugno}},
  \bibinfo{author}{\bibfnamefont{R.}~\bibnamefont{Saers}},
  \bibinfo{author}{\bibfnamefont{C.}~\bibnamefont{Fort}}, \bibnamefont{and}
  \bibinfo{author}{\bibfnamefont{M.}~\bibnamefont{Inguscio}},
  \bibinfo{journal}{Phys. Rev. Lett.} \textbf{\bibinfo{volume}{93}},
  \bibinfo{pages}{140406} (\bibinfo{year}{2004}).

\bibitem[{\citenamefont{Gemelke et~al.}(2005)\citenamefont{Gemelke, Sarajlic,
  Bidel, Hong, and Chu}}]{SChu2005PRL}
\bibinfo{author}{\bibfnamefont{N.}~\bibnamefont{Gemelke}},
  \bibinfo{author}{\bibfnamefont{E.}~\bibnamefont{Sarajlic}},
  \bibinfo{author}{\bibfnamefont{Y.}~\bibnamefont{Bidel}},
  \bibinfo{author}{\bibfnamefont{S.}~\bibnamefont{Hong}}, \bibnamefont{and}
  \bibinfo{author}{\bibfnamefont{S.}~\bibnamefont{Chu}},
  \bibinfo{journal}{Phys. Rev. Lett.} \textbf{\bibinfo{volume}{95}},
  \bibinfo{pages}{170404} (\bibinfo{year}{2005}).

\bibitem[{\citenamefont{Zhang et~al.}(2005{\natexlab{a}})\citenamefont{Zhang,
  Zhou, Chang, Chapman, and You}}]{You2005PRL}
\bibinfo{author}{\bibfnamefont{W.}~\bibnamefont{Zhang}},
  \bibinfo{author}{\bibfnamefont{D.~L.} \bibnamefont{Zhou}},
  \bibinfo{author}{\bibfnamefont{M.-S.} \bibnamefont{Chang}},
  \bibinfo{author}{\bibfnamefont{M.~S.} \bibnamefont{Chapman}},
  \bibnamefont{and} \bibinfo{author}{\bibfnamefont{L.}~\bibnamefont{You}},
  \bibinfo{journal}{Phys. Rev. Lett.} \textbf{\bibinfo{volume}{95}},
  \bibinfo{pages}{180403} (\bibinfo{year}{2005}{\natexlab{a}}).

\bibitem[{\citenamefont{Bloch}(2005)}]{InstabilityBEC}
\bibinfo{author}{\bibfnamefont{I.}~\bibnamefont{Bloch}}, \bibinfo{journal}{Nat
  Phys} \textbf{\bibinfo{volume}{1}}, \bibinfo{pages}{23 }
  (\bibinfo{year}{2005}), ISSN \bibinfo{issn}{1745-2473}.

\bibitem[{\citenamefont{Lieb et~al.}(2000)\citenamefont{Lieb, Seiringer, and
  Yngvason}}]{Lieb2000PRA}
\bibinfo{author}{\bibfnamefont{E.~H.} \bibnamefont{Lieb}},
  \bibinfo{author}{\bibfnamefont{R.}~\bibnamefont{Seiringer}},
  \bibnamefont{and} \bibinfo{author}{\bibfnamefont{J.}~\bibnamefont{Yngvason}},
  \bibinfo{journal}{Phys. Rev. A} \textbf{\bibinfo{volume}{61}},
  \bibinfo{pages}{043602} (\bibinfo{year}{2000}).

\bibitem[{\citenamefont{Erd\ifmmode~\mbox{\H{o}}\else \H{o}\fi{}s
  et~al.}(2007)\citenamefont{Erd\ifmmode~\mbox{\H{o}}\else \H{o}\fi{}s,
  Schlein, and Yau}}]{DeriveGPE}
\bibinfo{author}{\bibfnamefont{L.}~\bibnamefont{Erd\ifmmode~\mbox{\H{o}}\else
  \H{o}\fi{}s}}, \bibinfo{author}{\bibfnamefont{B.}~\bibnamefont{Schlein}},
  \bibnamefont{and} \bibinfo{author}{\bibfnamefont{H.-T.} \bibnamefont{Yau}},
  \bibinfo{journal}{Phys. Rev. Lett.} \textbf{\bibinfo{volume}{98}},
  \bibinfo{pages}{040404} (\bibinfo{year}{2007}).

\bibitem[{\citenamefont{Berman and Zaslavsky}(1978)}]{Berman1978}
\bibinfo{author}{\bibfnamefont{G.}~\bibnamefont{Berman}} \bibnamefont{and}
  \bibinfo{author}{\bibfnamefont{G.}~\bibnamefont{Zaslavsky}},
  \bibinfo{journal}{Physica A} \textbf{\bibinfo{volume}{91}},
  \bibinfo{pages}{450 } (\bibinfo{year}{1978}).

\bibitem[{\citenamefont{Silvestrov and Beenakker}(2002)}]{Ehrenfest}
\bibinfo{author}{\bibfnamefont{P.~G.} \bibnamefont{Silvestrov}}
  \bibnamefont{and} \bibinfo{author}{\bibfnamefont{C.~W.~J.}
  \bibnamefont{Beenakker}}, \bibinfo{journal}{Phys. Rev. E}
  \textbf{\bibinfo{volume}{65}}, \bibinfo{pages}{035208}
  (\bibinfo{year}{2002}).

\bibitem[{\citenamefont{Yaffe}(1982)}]{Yaffe}
\bibinfo{author}{\bibfnamefont{L.~G.} \bibnamefont{Yaffe}},
  \bibinfo{journal}{Rev. Mod. Phys.} \textbf{\bibinfo{volume}{54}},
  \bibinfo{pages}{407} (\bibinfo{year}{1982}).

\bibitem[{\citenamefont{Nemoto et~al.}(2000)\citenamefont{Nemoto, Holmes,
  Milburn, and Munro}}]{triplewell}
\bibinfo{author}{\bibfnamefont{K.}~\bibnamefont{Nemoto}},
  \bibinfo{author}{\bibfnamefont{C.~A.} \bibnamefont{Holmes}},
  \bibinfo{author}{\bibfnamefont{G.~J.} \bibnamefont{Milburn}},
  \bibnamefont{and} \bibinfo{author}{\bibfnamefont{W.~J.} \bibnamefont{Munro}},
  \bibinfo{journal}{Phys. Rev. A} \textbf{\bibinfo{volume}{63}},
  \bibinfo{pages}{013604} (\bibinfo{year}{2000}).

\bibitem[{\citenamefont{Franzosi and Penna}(2001)}]{triplewell2}
\bibinfo{author}{\bibfnamefont{R.}~\bibnamefont{Franzosi}} \bibnamefont{and}
  \bibinfo{author}{\bibfnamefont{V.}~\bibnamefont{Penna}},
  \bibinfo{journal}{Phys. Rev. A} \textbf{\bibinfo{volume}{65}},
  \bibinfo{pages}{013601} (\bibinfo{year}{2001}).

\bibitem[{\citenamefont{Franzosi and Penna}(2003)}]{Penna2003PRE}
\bibinfo{author}{\bibfnamefont{R.}~\bibnamefont{Franzosi}} \bibnamefont{and}
  \bibinfo{author}{\bibfnamefont{V.}~\bibnamefont{Penna}},
  \bibinfo{journal}{Phys. Rev. E} \textbf{\bibinfo{volume}{67}},
  \bibinfo{pages}{046227} (\bibinfo{year}{2003}).

\bibitem[{\citenamefont{Liu et~al.}(2007)\citenamefont{Liu, Fu, Yang, and
  Liu}}]{Liu2007PRA}
\bibinfo{author}{\bibfnamefont{B.}~\bibnamefont{Liu}},
  \bibinfo{author}{\bibfnamefont{L.-B.} \bibnamefont{Fu}},
  \bibinfo{author}{\bibfnamefont{S.-P.} \bibnamefont{Yang}}, \bibnamefont{and}
  \bibinfo{author}{\bibfnamefont{J.}~\bibnamefont{Liu}},
  \bibinfo{journal}{Phys. Rev. A} \textbf{\bibinfo{volume}{75}},
  \bibinfo{pages}{033601} (\bibinfo{year}{2007}).

\bibitem[{\citenamefont{Guo et~al.}(2014)\citenamefont{Guo, Chen, and
  Wu}}]{GuoWu}
\bibinfo{author}{\bibfnamefont{Q.}~\bibnamefont{Guo}},
  \bibinfo{author}{\bibfnamefont{X.}~\bibnamefont{Chen}}, \bibnamefont{and}
  \bibinfo{author}{\bibfnamefont{B.}~\bibnamefont{Wu}}, \bibinfo{journal}{Opt.
  Express} \textbf{\bibinfo{volume}{22}}, \bibinfo{pages}{19219}
  (\bibinfo{year}{2014}).

\bibitem[{\citenamefont{Luo et~al.}(2007)\citenamefont{Luo, Xie, and
  Wu}}]{TunnelWu}
\bibinfo{author}{\bibfnamefont{X.}~\bibnamefont{Luo}},
  \bibinfo{author}{\bibfnamefont{Q.}~\bibnamefont{Xie}}, \bibnamefont{and}
  \bibinfo{author}{\bibfnamefont{B.}~\bibnamefont{Wu}}, \bibinfo{journal}{Phys.
  Rev. A} \textbf{\bibinfo{volume}{76}}, \bibinfo{pages}{051802}
  (\bibinfo{year}{2007}).

\bibitem[{\citenamefont{Habib et~al.}(1998)\citenamefont{Habib, Shizume, and
  Zurek}}]{Zurek}
\bibinfo{author}{\bibfnamefont{S.}~\bibnamefont{Habib}},
  \bibinfo{author}{\bibfnamefont{K.}~\bibnamefont{Shizume}}, \bibnamefont{and}
  \bibinfo{author}{\bibfnamefont{W.~H.} \bibnamefont{Zurek}},
  \bibinfo{journal}{Phys. Rev. Lett.} \textbf{\bibinfo{volume}{80}},
  \bibinfo{pages}{4361} (\bibinfo{year}{1998}).

\bibitem{Anglin2001}A. Vardi and J.R. Anglin, 
Phys. Rev. Lett. {\bf 86}, 568 (2001).


\bibitem[{\citenamefont{Gertjerenken et~al.}(2010)\citenamefont{Gertjerenken,
  Arlinghaus, Teichmann, and Weiss}}]{ChristophPRA}
\bibinfo{author}{\bibfnamefont{B.}~\bibnamefont{Gertjerenken}},
  \bibinfo{author}{\bibfnamefont{S.}~\bibnamefont{Arlinghaus}},
  \bibinfo{author}{\bibfnamefont{N.}~\bibnamefont{Teichmann}},
  \bibnamefont{and} \bibinfo{author}{\bibfnamefont{C.}~\bibnamefont{Weiss}},
  \bibinfo{journal}{Phys. Rev. A} \textbf{\bibinfo{volume}{82}},
  \bibinfo{pages}{023620} (\bibinfo{year}{2010}).

\bibitem[{\citenamefont{Weiss and Teichmann}(2008)}]{ChristophPRL}
\bibinfo{author}{\bibfnamefont{C.}~\bibnamefont{Weiss}} \bibnamefont{and}
  \bibinfo{author}{\bibfnamefont{N.}~\bibnamefont{Teichmann}},
  \bibinfo{journal}{Phys. Rev. Lett.} \textbf{\bibinfo{volume}{100}},
  \bibinfo{pages}{140408} (\bibinfo{year}{2008}).
  
\bibitem[{\citenamefont{B\ifmmode~\check{r}\else \v{r}\fi{}ezinov\'a
  et~al.}(2012)\citenamefont{B\ifmmode~\check{r}\else \v{r}\fi{}ezinov\'a,
  Lode, Streltsov, Alon, Cederbaum, and Burgd\"orfer}}]{WaveChaosPRA}
\bibinfo{author}{\bibfnamefont{I.}~\bibnamefont{B\ifmmode~\check{r}\else
  \v{r}\fi{}ezinov\'a}}, \bibinfo{author}{\bibfnamefont{A.~U.~J.}
  \bibnamefont{Lode}}, \bibinfo{author}{\bibfnamefont{A.~I.}
  \bibnamefont{Streltsov}}, \bibinfo{author}{\bibfnamefont{O.~E.}
  \bibnamefont{Alon}}, \bibinfo{author}{\bibfnamefont{L.~S.}
  \bibnamefont{Cederbaum}}, \bibnamefont{and}
  \bibinfo{author}{\bibfnamefont{J.}~\bibnamefont{Burgd\"orfer}},
  \bibinfo{journal}{Phys. Rev. A} \textbf{\bibinfo{volume}{86}},
  \bibinfo{pages}{013630} (\bibinfo{year}{2012}).


\bibitem[{\citenamefont{Buonsante et~al.}(2011)\citenamefont{Buonsante, Penna,
  and Vezzani}}]{PennaPRA}
\bibinfo{author}{\bibfnamefont{P.}~\bibnamefont{Buonsante}},
  \bibinfo{author}{\bibfnamefont{V.}~\bibnamefont{Penna}}, \bibnamefont{and}
  \bibinfo{author}{\bibfnamefont{A.}~\bibnamefont{Vezzani}},
  \bibinfo{journal}{Phys. Rev. A} \textbf{\bibinfo{volume}{84}},
  \bibinfo{pages}{061601} (\bibinfo{year}{2011}).

\bibitem[{\citenamefont{Buonsante et~al.}(2012)\citenamefont{Buonsante,
  Burioni, Vescovi, and Vezzani}}]{PennaPRA2}
\bibinfo{author}{\bibfnamefont{P.}~\bibnamefont{Buonsante}},
  \bibinfo{author}{\bibfnamefont{R.}~\bibnamefont{Burioni}},
  \bibinfo{author}{\bibfnamefont{E.}~\bibnamefont{Vescovi}}, \bibnamefont{and}
  \bibinfo{author}{\bibfnamefont{A.}~\bibnamefont{Vezzani}},
  \bibinfo{journal}{Phys. Rev. A} \textbf{\bibinfo{volume}{85}},
  \bibinfo{pages}{043625} (\bibinfo{year}{2012}).

\bibitem[{\citenamefont{Karkuszewski et~al.}(2002)\citenamefont{Karkuszewski,
  Zakrzewski, and Zurek}}]{ZurekPRA}
\bibinfo{author}{\bibfnamefont{Z.~P.} \bibnamefont{Karkuszewski}},
  \bibinfo{author}{\bibfnamefont{J.}~\bibnamefont{Zakrzewski}},
  \bibnamefont{and} \bibinfo{author}{\bibfnamefont{W.~H.} \bibnamefont{Zurek}},
  \bibinfo{journal}{Phys. Rev. A} \textbf{\bibinfo{volume}{65}},
  \bibinfo{pages}{042113} (\bibinfo{year}{2002}).

\bibitem[{\citenamefont{Chang et~al.}(2005)\citenamefont{Chang, Qin, Zhang,
  You, and Chapman}}]{Chapman2005NPhys}
\bibinfo{author}{\bibfnamefont{M.-S.} \bibnamefont{Chang}},
  \bibinfo{author}{\bibfnamefont{Q.}~\bibnamefont{Qin}},
  \bibinfo{author}{\bibfnamefont{W.}~\bibnamefont{Zhang}},
  \bibinfo{author}{\bibfnamefont{L.}~\bibnamefont{You}}, \bibnamefont{and}
  \bibinfo{author}{\bibfnamefont{M.~S.} \bibnamefont{Chapman}},
  \bibinfo{journal}{Nat Phys} \textbf{\bibinfo{volume}{1}},
  \bibinfo{pages}{111} (\bibinfo{year}{2005}).

\bibitem[{\citenamefont{Zhang et~al.}(2005{\natexlab{b}})\citenamefont{Zhang,
  Zhou, Chang, Chapman, and You}}]{YouLiPRA2005}
\bibinfo{author}{\bibfnamefont{W.}~\bibnamefont{Zhang}},
  \bibinfo{author}{\bibfnamefont{D.~L.} \bibnamefont{Zhou}},
  \bibinfo{author}{\bibfnamefont{M.-S.} \bibnamefont{Chang}},
  \bibinfo{author}{\bibfnamefont{M.~S.} \bibnamefont{Chapman}},
  \bibnamefont{and} \bibinfo{author}{\bibfnamefont{L.}~\bibnamefont{You}},
  \bibinfo{journal}{Phys. Rev. A} \textbf{\bibinfo{volume}{72}},
  \bibinfo{pages}{013602} (\bibinfo{year}{2005}{\natexlab{b}}).

\end{thebibliography}
\end{document}